\documentstyle[prd,aps]{revtex}
\def\Caption#1#2{\refstepcounter{figure}\label{#1}\noindent{Fig.~\thefigure:}#2\par\vspace*{12truept}}
\begin{document}
\author{S.~W.~Mansour and T.~K.~Kuo}
\address{Department of Physics, Purdue University, West Lafayette, Indiana 47907}
\title{Supernova neutrinos in the light of FCNC}
\date{}
\maketitle
\input epsf
\begin{abstract}
We study the effect of including flavor changing neutral currents (FCNC) in the analysis of the neutrino signal of a supernova burst. When we include the effect of the FCNC which are beyond the standard model (SM) in the study of the MSW resonant conversion, we obtain dramatic changes in the $\Delta m^2-\sin^2{2\theta} $ probability contours for neutrino detection. 
\end{abstract}
\section{Introduction}
It is well known that there is considerable discrepancy between the observed solar neutrino fluxes and their theoretical predictions based on the standard solar models. To solve the longstanding solar neutrino problem~\cite{a:1}, the most natural scenario is to include the effect of matter on neutrino oscillations (the MSW effect)~\cite{a:2}. 
The SM interaction between the neutrinos and the electrons in solar matter would enhance the $\nu_{e}\leftrightarrow\nu_{\mu,\tau}$ conversion and that could account for the apparent deficit in the solar neutrino flux. If indeed the MSW effect is the cause of the solar neutrino problem, a similar situation would exist for the neutrinos produced in supernova explosions, such as those detected from the 1987A supernova. The effect of the MSW effect in the light of the SM on the supernova neutrino signal has been studied extensively~\cite{a:3,a:4}.

Recently, there has been a lot of discussions on possible new neutrino interactions beyond those of the SM. If these interactions do exist, they would have to be included in the study of neutrino propagation in matter. Interest in this subject has heightened especially after the recent announcement of observing an excess of large-$Q^2$  $\textrm{e}^+\textrm{p}$ scattering events at HERA~\cite{a:5}, with a possible signature for leptoquark formation. Also, possible  Flavor Changing Neutral Currents (FCNC) and Flavor Diagonal Neutral Currents (FDNC) arise naturally in a number of extensions of the SM, such as the  Left-Right Symmetric Models~\cite{a:6} and Supersymmetric models with R-parity violation~\cite{a:7}. In fact, they have been recently discussed in the study of solar neutrinos~\cite{a:8,a:9,a:10,a:11}. 

In this paper, we study the effect of these new physics interactions on the supernova neutrino signal. It will be another test of these new models which covers a wider range of neutrino parameters compared to the solar neutrino analysis. We neglect the non-standard FDNC interactions as the current limits on them make their effect negligible if neutrinos have mass.
In section 2, we use the three-flavor formalism to derive the survival probability for neutrinos in the presence of FCNC beyond the SM. The three-flavor formalism is appropriate for describing supernova neutrinos because of the large range of densities in the supernova which the neutrinos must traverse. This means that, for all conceivable neutrino mass differences, both $\nu_e\leftrightarrow\nu_\mu$ and $\nu_e\leftrightarrow\nu_\tau$ conversions will go through the resonance regions~\cite{a:3}.  
In section 3, we show the results of our calculations for the suppression (enhancement) ratio of the neutrino events due to supernova bursts which can be observed by the SuperKamiokande~\cite{a:12} and the Sudbury Neutrino Observatory (SNO)~\cite{a:13} (already in the construction phase). The latter detector has the advantage of being sensitive to all neutrino flavors, not just $\nu_e$'s.
We use typical values for the initial fluxes of the different neutrino flavors produced during a supernova collapse based on theoretical models describing supernova collapse and which are in a good agreement with the 1987A event.
In our $\delta m^2-\sin^2\theta$ contours, we emphasize the case where we have only $\nu-d$ scattering, however, other new physics interactions can be easily included in the analysis.

\section{Neutrino FCNC Interactions}
The only relevant SM interaction which is effective in the MSW effect is the standard $W$-exchange charged current $\nu_e e \rightarrow \nu_e e$. However, in supersymmetric theories without R-parity and other non-SM models, new types of possible neutrino interactions with matter enter the picture. We can divide these interactions into two types: Flavor Changing Neutrino Currents (FCNC) and Flavor Diagonal Neutral Currents (FDNC). To be specific, we will present an analysis based on R-parity violating supersymmetric models. However, similar results as well as our parameterization of the effective interactions will follow had we chosen to use other theoretical models. In particular, we will be interested in the following lepton-number-violating contributions to the superpotential~\cite{a:8}:
$$\lambda_{ijk}L_iL_jE_k^c,\eqno(2.1a)$$
$$\lambda'_{ijk}L_iQ_jD_k^c,\eqno(2.1b)$$
where $L, Q, E^c, D^c$ are the usual lepton and quark SU(2) doublets and singlets, respectively, and $i,j,k$ are generation indices.
Because of the contraction of the SU(2) indices, the coefficients $\lambda_{ijk}$  should be antisymmetric under the exchange of $i$ and $j$. We will assume the coupling to be real to conserve CP. The superpotentials, Eqs. (2.1), lead to the Lagrangians
$${\cal{L}}_L=\lambda_{ijk}[\tilde{\nu}_L^i\bar{e}_R^k e_L^j+\tilde{e}_L^j\bar{e}_R^k\nu_L^i+(\tilde{e}_R^k)^*(\bar{\nu}^i_L)^c e_L^j - (i\leftrightarrow j)]+ \textrm{H.C.} 
\eqno(2.2)$$ and
\begin{eqnarray}
{\cal{L}}_Q&=&\lambda'_{ijk}[\tilde{\nu}_L^i\bar{d}_R^k d_L^j+\tilde{d}_L^j\bar{d}_R^k\nu_L^i+(\tilde{d}_R^k)^*(\bar{\nu}^i_L)^c d_L^j\nonumber\\
 & &\mbox{}-\tilde{e}_L^i\bar{d}_R^k u_L^j-\tilde{u}_L^j\bar{d}_R^ke_L^i-(\tilde{d}_R^k)^*(\bar{e}_L^i)^c u_L^j]+ \textrm{H.C.}
\eqnum{2.3}  
\end{eqnarray}
These interactions can give rise to $\nu-d$ and $\nu-e$ scattering, via squark and selectron exchange, respectively, if the couplings $\lambda_{ijk}$ and $\lambda'_{ijk}$ are nonzero. The current limits on $\lambda_{ijk}$ and $\lambda'_{ijk}$ can be determined from charged-current universality constraints, the branching ratios of lepton-number violating interactions, etc.
A detailed account of the limits on $\lambda_{ijk}$ and $\lambda'_{ijk}$ is given in ~\cite{a:14}.

To obtain the neutrino survival probability, we start with the three-flavor propagation equation for neutrinos (see Ref.~\cite{a:15}),
\begin{eqnarray}
i{d\over dx}\left(\matrix{\nu_e\cr \nu_\mu\cr \nu_\tau}\right)&=&{1\over 2E}M^2\left(\matrix{\nu_e\cr \nu_\mu\cr \nu_\tau}\right)\nonumber \\&=&{1\over 2E}\left[U\left(\matrix{m_1^2&0&0\cr 0&m_2^2&0\cr 0&0&m_3^2}\right)U^\dag + \left(\matrix{A&B&C\cr B&0&0\cr C&0&0}\right)\right]\left(\matrix{\nu_e\cr \nu_\mu\cr \nu_\tau}\right),
\eqnum{2.4}
\end{eqnarray} 
where $A\equiv 2E\sqrt{2}G_F N_e$ is the SM squared-induced mass, $E$ is the neutrino energy, $N_e$ is the electron density in the supernova core, while $U$ is the three-flavor mixing matrix given by:
\begin{eqnarray}
U &=&\exp\left({i\psi {\lambda }_7}\right)\Gamma\exp \left({i\varphi{\lambda }_ 5}\right)\exp \left({i\omega {\lambda }_ 2}\right)]\nonumber\\
&=& \left[{\matrix{1&0&0\cr 0&{C}_{\psi }&{S}_{\psi }\cr0&{-S}_{\psi }&{C}_{\psi }\cr}}\right]\left[{\matrix{1&0&0\cr 0&{e}^{i\delta }&0\cr 0&0&{e}^{-i\delta }\cr}}\right]\left[{\matrix{{C}_{\varphi }&0&{S}_{\varphi }\cr0&1&0\cr{-S}_{\varphi }&0&{C}_{\varphi }\cr}}\right]\left[{\matrix{{C}_{\omega }&{S}_{\omega}&0\cr {-S}_{\omega }&{C}_{\omega }&0\cr 0&0&1\cr}}\right],
\eqnum{2.5}
\end{eqnarray}
$\lambda_i$ are the Gell-Mann matrices while $\psi$, $\varphi$ and $\omega$ are the mixing angles corresponding to $\nu_\mu\leftrightarrow\nu_\tau$, $\nu_e\leftrightarrow\nu_\tau$ and $\nu_e\leftrightarrow\nu_\mu$ oscillations  respectively. For simplicity, we will set the CP violating phase, $\delta$, to zero in our calculation.
The new physics parameters $B$ and $C$ describe the additional FCNC contributions due to $\nu_e\leftrightarrow\nu_\mu$ and $\nu_e\leftrightarrow\nu_\tau$ interactions, respectively. They are given in terms of the couplings as
\begin{eqnarray}
B&\equiv&2E\sqrt{2}(G_N^e N_e+G_N^u N_u+G^d_N N_d) \nonumber\\
&=&2E\sqrt{2}G_FN_e\left[\epsilon_e+2\epsilon_u+\epsilon_d+(\epsilon_u+2\epsilon_d){N_n\over N_e}\right],
\eqnum{2.6a}
\end{eqnarray}
\begin{eqnarray}
C&\equiv&2E\sqrt{2}(G_N^{'e} N_e+G_N^{'u} N_u+G^{'d}_N N_d) \nonumber\\
&=&2E\sqrt{2}G_FN_e\left[\epsilon'_e+2\epsilon'_u+\epsilon'_d+(\epsilon'_u+2\epsilon'_d){N_n\over N_e}\right],
\eqnum{2.6b}
\end{eqnarray}
where $N_f$ denotes the number density of the fermion type $f$, $G_N^f$ and $G_N^{'f}$ are the effective couplings of the reaction $\nu_e f \rightarrow \nu_\mu f$ and $\nu_e f \rightarrow \nu_\tau f$, respectively; while $\epsilon_f(\epsilon'_f)\equiv G_N^f/G_F(G_N^{'f}/G_F)$. Note that we assumed an equal number density of protons and electrons inside the supernova core.
The order of the new physics corrections can be at most $10^{-2}$ and this is due to the current bounds from rare decays, atomic parity violation, etc.  The largest corrections would be due to transitions between the first and third generations~\cite{a:10}. The relevant parameters in terms of the $b$ squark mass, $m_{\tilde{b}}$, are:
$$\epsilon_d={\lambda'_{231}\lambda'_{131}\over 4\sqrt{2}G_F m_{\tilde{b}}^2},$$
$$\epsilon'_d={\lambda'_{331}\lambda'_{131}\over 4\sqrt{2}G_F m_{\tilde{b}}^2}.\eqno(2.7)$$
They are currently bounded from above at the order of $10^{-2}$.

To find the effective mixing angles in matter, we diagonalize $M^2$. We can rotate the flavor basis by the factor $\exp(-i\varphi\lambda_5)\exp(-i\psi\lambda_7)$ and find the eigenvalues of the transformed mass matrix 
\begin{eqnarray}
\tilde{M}^2&\equiv& \exp(-i\varphi \lambda_5)\exp(-i\psi \lambda_7)M^2\exp(i\psi \lambda_7)\exp(i\varphi\lambda_5) \nonumber\\
&=&{1\over 2}\left(\matrix{\Sigma-\Delta C_{2\omega}+2AC_\varphi^2+d_0&\Delta S_{2\omega}+d_1&AS_{2\varphi}+d_2\cr \Delta S_{2\omega}+d_1&\Sigma+\Delta C_{2\omega}&d_3\cr A S_{2\varphi}+d_2&d_3&2(m_3^2+AS_\varphi^2)-d_0}\right),
\eqnum{2.8}
\end{eqnarray} 
where 
\begin{tabbing}
\hspace{2.5 true in}\=  \kill
\>$\Sigma=m_1^2+m_2^2$, $\Delta=m_2^2-m_1^2$,\\ 
\>$d_0=-2S_{2\varphi}(BS_\psi+CC_\psi)$,\\ 
\>$d_1=2C_\varphi(BC_\psi-CS_\psi)$,\\
\>$d_2=2C_{2\varphi}(BS_\psi+CC_\psi)$,\\
\>$d_3=2S_\varphi(BC_\psi-CS_\psi)$. 
\end{tabbing}

We are using the same notation of Ref.~\cite{a:3} with $C_\varphi\equiv\cos\varphi$, $S_\varphi\equiv\sin\varphi$, etc.
To zeroth order in $A\sin 2\varphi$, the first matter-enhanced mixing angle due to the first resonance region is given by
$$\tan 2\omega_m={\Delta\sin 2\omega+D_l\over -A\cos^2\varphi+\Delta\cos 2\omega},\eqno(2.9)$$
with the new physics correction $D_l\equiv d_1=2(B \cos\varphi\cos\psi - C\sin\psi\cos\varphi)$.
The lower resonance occurs at $A\cos^2\varphi=\Delta \cos 2\omega$.
 
We now turn to the higher resonance region, we rotate the flavor basis by $\exp(-i\psi\lambda_7)$, we obtain for $\tilde{M}^2$:
$$\tilde{M}^2={1\over 2}\left(\matrix{2(A+m_3^2 S_\varphi^2)+\Lambda C_\varphi^2&\Delta S_{2\omega}C_\varphi+b_0&(m_3^2-\Lambda/2)S_{2\varphi}+b_1\cr
\Delta S_{2\omega}C_{\varphi}+b_0&\Sigma+\Delta C_{2\omega}&-\Delta S_\varphi S_{2\omega}\cr (m_3^2-\Lambda/2)S_{2\varphi}+b_1&-\Delta S_\varphi S_{2\omega}&\Lambda S_\varphi^2+2m_3^2C_\varphi^2}\right),\eqno(2.10)$$
where $\Lambda=\Sigma-\Delta C_{2\omega}$, $b_0=2(BC_\psi-CS_\psi)$ and $b_1=2(BS_\psi+CC_\psi)$.  
The second matter-enhanced mixing angle is (to zeroth order in $\Delta\sin 2\omega/2 m_3^2$)
 $$\tan 2\varphi_m={m_3^2\sin 2\varphi+D_h\over -A+m_3^2\cos 2\varphi},\eqno(2.11)$$
where $D_h\equiv b_1=2(C\cos\psi+B\sin\psi)$ and the higher resonance is at $A=m_3^2\cos 2\varphi$.
Note that our expressions for the mixing angles lead to the SM three-flavor solutions~\cite{a:15} when $D_{l,h}\rightarrow 0$.
If we assume a mass hierarchy, where $m_3^2\gg m_2^2 \gg m_1^2$, we can separate the two resonance regions, each resembling a two-flavor resonance and the three-flavor survival probability can be written as
 $$P^3(\nu_e \rightarrow \nu_e)=P^l(\nu_e\rightarrow\nu_e)P^h(\nu_e\rightarrow\nu_e),\eqno(2.12)$$
where $P_l$ and $P_h$ are given by ~\cite{a:16}:
$$P_l(\nu_e\rightarrow\nu_e)\equiv {1\over 2}+\left({1\over 2}-\Theta(A\cos^2\varphi-\Delta \cos 2\omega)P_c(\gamma,\omega)\right)\cos 2\omega \cos 2\omega_m\eqno(2.13a)$$
and $$P_h(\nu_e\rightarrow\nu_e)\equiv {1\over 2}+\left({1\over 2}-\Theta(A-m_3^2 \cos 2\varphi)P_c(\gamma,\varphi)\right)\cos 2\varphi\cos2\varphi_m.\eqno(2.13b)$$
The level-crossing probability, $P_c$, is given by
$$P_c\equiv|\left<\nu_2(x_+)|\nu_1(x_-)\right>|=\exp\left[-{\pi\over2}\gamma F(\theta)\right];\eqno(2.14)$$ 
$\gamma$ being the adiabaticity parameter and $F(\theta)$ is a function
dependent on the density profile which reduces to unity for a linear density function.
Note that the adiabaticity parameter, $\gamma$, is modified from its MSW form, it is given by~\cite{a:11} 
$$\gamma=\gamma_{SM}\left|1+\epsilon_0\cot 2 \theta\right|^2,\eqno(2.15)$$
where $\gamma_{SM}={\delta \sin^2 2\theta \over 2E\cos 2\theta (dN_e/dx)/N_e|_{res}}$, while $\delta=(\Delta\equiv m_2^2-m_1^2,m_3^2)$ and $\theta=(\omega,\varphi)$ for the lower and higher resonances respectively.
We used the two-flavor result as both $P_l$ and $P_h$ have the two-flavor form with just a redefinition of variables.
The parameter $\epsilon_0$ is given by
$$\epsilon_0\equiv\epsilon_e+2\epsilon_u+\epsilon_d+(\epsilon_u+2\epsilon_d){N_n\over N_e}|_{resonance}.\eqno(2.16)$$
\section{Supernova Neutrinos}
We will now apply our results to neutrinos emitted in a supernova explosion. As neutrinos come out of the core of the supernova, they would traverse through a medium whose density decreases to zero from a maximum value of about $10^{12}$ gm/cm$^3$. This means that the resonance condition, $A=\Delta \cos 2\theta$, will be satisfied for all conceivable neutrino masses. Thus, unlike the case of the solar neutrinos, supernova neutrinos will go through two resonance regions as they stream out from the core and, hence, the three neutrino formalism presented in the previous section is necessary to describe them.

During a type II supernova burst, we have two types of neutrino emission processes, neutronization and thermal processes. The former is due to electron capture process, $e^-+p\rightarrow n+\nu_e,$ while the latter occurs during the implosion of the supernova inner layers leading to the thermal production of neutrinos of all flavors in equal amounts. We will assume a Fermi-Dirac distribution for both the neutronization and thermal fluxes, $F_\alpha^{0n}$ and $F_\alpha^{0t}$,  with the following initial temperatures and luminosities~\cite{a:17}
$$L_\alpha^t=L_{\bar{\alpha}}^t,\hspace{0.5in}T_e^t=T_{\bar e}^t=3 \textrm{MeV},$$
$$L_\mu^t=L_\tau^t\equiv L_x^t,\hspace{0.5in} T_x^t=T_{\bar{x}}^t=6 \textrm{MeV},$$
$$L_e^t=2L_x^t,\hspace{0.5in} T_e^n=3 \textrm{MeV},\eqno(3.1)$$  
where the superscripts $t$ and $n$ stand for thermal and neutronization,  while the subscripts $e$, $x$ and $\alpha$ denote the electron flavor, muon or tau flavors, and any flavor, respectively. Of course, there are uncertainties related to these values due to the rarity of supernova events and the complication of theoretical calculations and there are now more detailed numerical calculations on the time evolution of supernova. However, our approximate approach seems reasonable since there are very few neutrino events available at the present and the results shown are quite insensitive to the model dependent quantities for a supernova.
In addition, we will need the density profile of the core and mantle after the core collapse, it is given approximately by~\cite{a:18}
$$\rho(r)=C/r^3,\hspace{0.5in}10^{-5}<\rho<10^{12}\textrm{g/cm}^3,\eqno(3.2)$$
with $C$ varying weakly with $r$ over the range $1<C/10^{31}\textrm{g}<15$.
We assume the ratio $N_e/N_n\sim 0.4 ~\cite{a:19}$.

Knowing the survival probability, $P^3(\nu_e \rightarrow \nu_e)$, we can plot the ratio of the number of events in a detector with oscillations present to the number of events without oscillations. For a light-water Cerenkov detector there are two kinds of neutrino events: inverse beta decay which is isotropic and the $\nu-e$ scattering which is extremely directional and occurs for all flavors. These two classes of events combined with the two supernova emission processes leads to three relevant ratios to calculate. We show our plots for directional events from neutronization, the relevant ratio is given by~\cite{a:3}
$$R_{SK}={\int{dE_\nu F_e^{0n}\left[P\int{\epsilon d\sigma(\nu_e+e)}+(1-P)\int{\epsilon d\sigma(\nu_x+e)}\right]}\over{\int{dE_\nu F_e^{0n}\int{\epsilon d\sigma(\nu_e+e)}}}},\eqno(3.3)$$
Here $\int{\epsilon d\sigma}$ represents an integral over electron energies (from 0 to $E_\nu$) of the differential cross section times $\epsilon$, the
energy-dependent detector efficiency; $\nu_x$ represents $\nu_\mu$ or $\nu_\tau$. $P$ is the survival probability of electron neutrinos given by the three-flavor expression, Eq. (2.12), which depends on four independent parameters $U_{e2}$, $U_{e3}$, $\delta$ and $m_3$; so, in order  to make a 2-$d$ plot, we will have to constrain two of the four parameters.
In the case of a $\textrm{D}_2\textrm{O}$ detector, of which the SNO detector is an example, we will have the following important reactions with the deuterium nucleus in addition to the usual $e-\nu$ scattering: 
$$\nu(\bar{\nu})+d \rightarrow n+p+\nu(\bar{\nu}),\eqno(3.4a)$$ and$$\nu_e(\bar{\nu}_e)+d \rightarrow p(n) + p(n) + e^-(e^+).\eqno(3.4b)$$
Of course, the cross sections for these processes are very difficult to obtain analytically and we will use fit functions in our calculations to represent the cross sections, $\sigma(\nu+d)$, given in~\cite{a:20}.
The ratio of events with oscillations to those without will be equal to unity for the first type of event as it is equally sensitive to all flavors of neutrinos while the corresponding ratio for the second type is given by
$$R_{SNO}={\int{dE_\nu \epsilon\left[P F_e^{0t}+\left(1-P\right)F_x^{0t}\right]\sigma(\nu_e+d)}\over\int{dE_\nu \epsilon F_e^{0t} \sigma(\nu_e+d)}},\eqno(3.5)$$
 where again $\epsilon$ is the efficiency of our detector.
We assume in our calculations $\epsilon$ to be the same for the SuperKamiokande and SNO detectors. The efficiency of detection depends on the electron energy as~\cite{a:21}
$$\epsilon(E)=1-\exp\left[-\left((E-4.2)/4\right)^2\right],\eqno(3.6)$$
where $E$ is measured in MeV.

Figs.~(2) and (3) show the contour plots for the SuperKamiokande detector and the SNO detector. Plots for different values of $\epsilon_d\equiv G_N^d/G_F$,  the ratio between the new physics coupling to the SM coupling, are shown. Note that we get a similar behavior in the case of the study of solar neutrinos (see Ref.~\cite{a:11}). 
We use the following constraints: 
$m_1=0, (m_3/m_2)=250$ and $(|U_{e2}|^2/|U_{e3}|^2)=100$. We neglect the earth regeneration effects in our plots. 
In Fig.~(2), we have the expected SM behavior with the upper part of the contour corresponding to the adiabatic solution while the lower left side of the triangles correponding to the non-adiabatic solution. Note the two resonance regions correponding to the conversions between different flavors. The isocontours for the two experiments are the same since we assumed the same efficiency for the two detectors which is only an approximation. 
Fig.~(3) shows the effect of including non-SM couplings. At $\epsilon_d=\pm 10^{-2}$, the shape of the contours is significantly different from the SM case. We have a central region occuring at all values of the mixing angle where the event ratio is maximally different from 1.0 (0.2 in the case of SK and 2.3 in the case of SNO). Note that we have an enhancement in the number of observed events in the case of the SNO detector. At $\epsilon_d=10^{-4}$, we regain the adiabatic solution at large $\delta m^2$, also the central region divides into two regions. For negative couplings, we have an additional bump due to the divergence occuring at $\tan 2\theta=-2\epsilon_d $ (leading to $\gamma=\gamma_{SM}$). Note that for large values of the mixing angle, the effect of the new physics coupling becomes negligible and the large angle solution in the far right part of the plots becomes the same as the SM solution. 
Finally as we reach very small values of the couplings, $\epsilon_d\sim 10^{-6}$, we start to regain the SM solution.   

We should note that the parameters ($\delta m^2,\sin^2 2\theta$) are already bounded by other experiments: CHOOZ~\cite{a:22}, SuperKamiokande studies of solar~\cite{a:23} and atmospheric~\cite{a:24} neutrinos, etc.
The current bounds can be summarized as:
\begin{itemize}
\item Small angle solution for solar neutrinos based on the MSW effect:
$$3\times 10^{-6} \textrm{eV}^2 \lesssim \Delta m^2\lesssim 1\times 10^{-5} \textrm{eV}^2,$$
$$4\times 10^{-3} \lesssim \sin^2 2\theta\lesssim 1.3\times 10^{-2}.\eqno(3.7a)$$
\item Large angle solution for solar neutrinos based on the MSW effect:
$$6\times 10^{-6} \textrm{eV}^2 \lesssim \Delta m^2\lesssim 1\times 10^{-4} \textrm{eV}^2,$$
$$0.5 \lesssim \sin^2 2\theta\lesssim 0.9. \eqno(3.7b)$$
\item Vacuum oscillation solution for solar neutrinos:
$$5\times 10^{-11} \textrm{eV}^2 \lesssim \Delta m^2\lesssim 1\times 10^{-10} \textrm{eV}^2,$$ 
$$0.7 \lesssim \sin^2 2\theta\lesssim 1. \eqno(3.7c)$$
\item Solution for atmospheric neutrinos:
$$3\times 10^{-4} \textrm{eV}^2 \lesssim \Delta m^2\lesssim 7\times 10^{-3} \textrm{eV}^2,$$
$$0.8\lesssim \sin^2 2\theta\lesssim 1 .\eqno(3.7d)$$ 
\end{itemize}
\section{Conclusion}
Supernova explosions provide us with a very useful laboratory for testing the ideas of neutrino oscillations in matter as after their production in a supernova core, neutrinos travel through a wide range of densities not available anywhere else.
In this paper, we include non-standard FCNC neutrino interactions into the treatment of neutrino propagation in a supernova.  We may parametrize these effects as four-Fermi effective interactions ($\sim \epsilon G_F \nu \bar{\nu} f\bar{f})$.
Depending on the process, stringent bounds already exist on $\epsilon$, typically of the order of $10^{-2}$ or less. However, it is found that the changes in the rates of neutrino signals are appreciable even for $\epsilon$ as small as $10^{-6}-10^{-4}$. We showed numerical results in connection with the SuperKamiokande and the SNO detectors. It should be noted that the dependence of $\epsilon$ is intertwined with that of the neutrino mixing angles. Our results are therefore especially useful when combined with other measures which determine the bounds on mixing angles or when more definitive information on $\epsilon$ are available from direct measurements~\cite{a:5}. In conclusion, observing neutrino signals from supernovae would be another test for non-standard models which include FCNC interactions, such as the R-parity violating SUSY models. It may impose new constraints on the non-standard neutrino couplings or, more interestingly, signal new experimental evidence for these models.
This treatment is complementary to other studies of FCNC interactions in supernova where the bounds on these interactions are inferred from their effect on electron antineutrino spectrum and r-processes~\cite{a:25}.
\acknowledgements
S.~M.~was supported in part by a research grant from the Purdue Research Foundation, grant no.~690 1396-2714. T.~K.~is supported by a Department of Energy grant no.~DE FG02-91ER 40681. We would like to thank Sven Bergmann for his useful comments.

\pagebreak
\pagestyle{empty}
\section*{List of Figures}
\vspace{1 true in}
\Caption{fig1}{ The survival probability $P^3(\nu_e\rightarrow\nu_e)$ versus the neutrino energy $E$ for the case of $\epsilon_d=0$ (solid curve), $\epsilon_d= 0.01$ (dashed curve) and $\epsilon_d=-0.01$ (dotted curve).}
\Caption{fig2}{ Contour plot for the SuperKamiokande and SNO detector for $\epsilon_u=\epsilon_d=\epsilon_e=0$.}
\Caption{fig3}{ Contour plots for the Kamiokande and SNO detectors for $\epsilon_d=\pm 10^{-2}, \pm 10^{-4}$ and \mbox{$\pm 10^{-6}$.}}
\clearpage
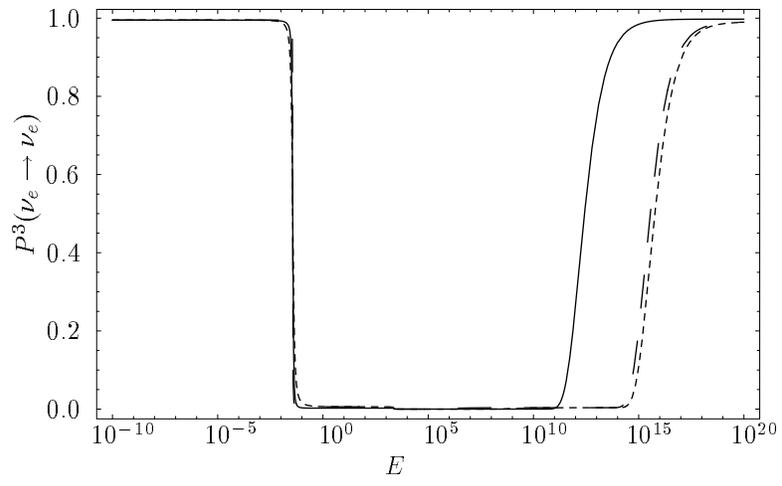
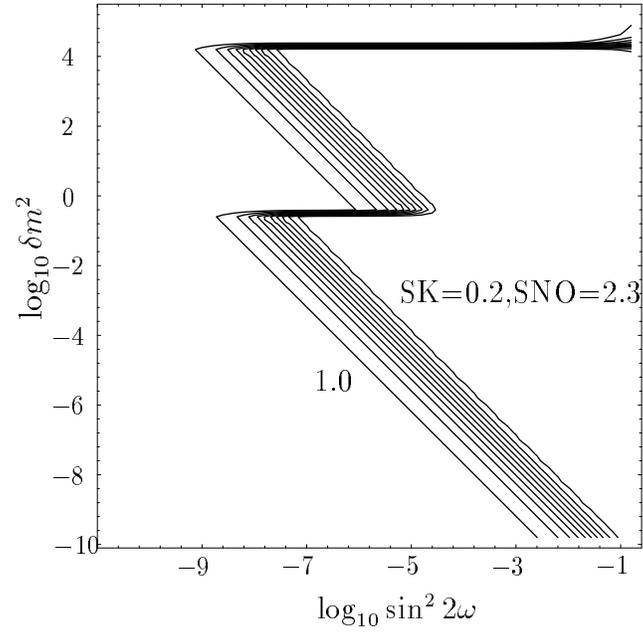
\end{document}